\documentstyle[aps,epsfig]{revtex}

\begin{document}

\title{Quantum corrections for pion correlations involving resonance decays}
\draft
\author{Sen Cheng and Scott Pratt}
\date{\today}
\address{Department of Physics and Astronomy and\\
  National Superconducting Cyclotron Laboratory,\\
  Michigan State University, East Lansing, Michigan 48824-1321}

\maketitle

\bigskip 
\begin{abstract}
  A method is presented to include quantum corrections into the calculation of
  two-pion correlations for the case where particles originate from resonance
  decays. The technique uses classical information regarding the space-time
  points at which resonances are created. By evaluating a simple thermal
  model, the method is compared to semiclassical techniques that assume
  exponential decaying resonances moving along classical
  trajectories. Significant improvements are noted when the resonance widths
  are broad as compared to the temperature.
\end{abstract}


\section{Introduction}

Analyses of two-boson correlations have provided intangible information
regarding the space-time development of hadronic reactions
\cite{Heinz:1999rw,Wiedemann:1999qn}. Pions, kaons, and photons have all been
exploited for their bosonic nature which results in a positive correlation at
small relative momentum. Numerous other correlations, involving
nucleons\cite{pp_gelbke,pp_ags,nn_gaff,nn_colonna} or light
nuclei\cite{imf_lacey,imf_desouza} that are correlated due to the strong or
Coulomb interaction as well as identical-particle statistics, have also been
analyzed and have given further information regarding collision
dynamics. Source sizes and time scales have been extracted from collisions
covering a wide assortment of reactions, from heavy-ion collisions at a few MeV
per nucleon, where time scales of thousands of fm/$c$ have been determined, to
$e^+e^-\rightarrow$ jets, where lifetimes of a fraction of a fm/$c$ have been
observed \cite{epluseminus_correlations}.

The comparison of theoretically predicted correlation functions with
experimental results provides an important test of the dynamical properties of
reaction models. Most models provide semiclassical information about the
source function $S({\bf p},x)$, the probability of emitting a particle of
momentum ${\bf p}$ from the space-time point $x$. By convoluting the source
functions for particles of momenta ${\bf p}_1$ and ${\bf p}_2$ with the squared
relative wave function $|\phi_{\bf q}(x_1-x_2)|^2$, one is able to predict the
correlation function $C({\bf p}_1,{\bf p}_2)$.  Source functions are usually
obtained from semiclassical simulations, where the source points are
associated with the last point of interaction \cite{pratt_qm93}. Particles from
resonances are usually assumed to be emitted according to an exponential decay
law, with the characteristic time usually chosen to be independent of the
energy of the resonance. Quantum considerations have been explored by Lednicky
and Progulova\cite{Lednicky:1992me} and by Bertsch, Danielewicz and Herrmann
\cite{Bertsch:1993nx}.

In this study, we pursue two goals. First, we wish to quantify the importance
of quantum treatments by comparing to semiclassical forms for a simple thermal
model. Although the formalism we present is not much different from that
presented previously in the literature \cite{Lednicky:1992me,Bertsch:1993nx},
the differences with semiclassical treatments have not been quantitatively
documented. We find that quantum corrections become important when kinematics
constrain the resonances to be off shell. Second, we propose an alteration to
the methods for extracting correlations from classical simulations to better
account for quantum effects. We show that a simple modification of the
semiclassical treatment can account for the quantum corrections by
incorporating information regarding the off-shell energy of the decaying
resonance. In this study we neglect any interaction between the particles aside
from the constraints imposed by symmetrization.

In this paper, we will first briefly review correlation functions for direct sources, which
will provide the foundation for correlations from resonant sources in the
following section. Section ~\ref{sec:narrowresonance} contains a calculation
for 
a simple Breit-Wigner resonance to demonstrate the importance of quantum
treatments. Modified correlation weights are
presented in Sec.~\ref{sec:weights}  as a means to better calculate correlations from classical
simulations when resonant decays are involved. Finally, we will compare the
various methods mentioned in this paper by considering a simple thermal model
in Sec. \ref{sec:results}.

\section{Review: Correlations from direct sources}
\label{sec:review_direct}

The two-particle correlation function is usually
\label{Heinz:1999rw,Wiedemann:1999qn} defined as
\begin{equation}
  \label{eq:correl_def}
  C({\bf p}_1,{\bf p}_2)\equiv
  \frac{ dN^{(2)}/(d^3p_1d^3p_2)}
  { (dN^{(1)}/d^3p_1)~ (dN^{(1)}/d^3p_2)}
\end{equation}
Before describing the two-particle probability, we first derive an expression
for the one-particle probability, which also introduces the definition of the
source function. One-particle probabilities can be determined by matrix
elements $T_f(x)$ where $f$ describes the remainder of the system, and $x$ is
the point at which the pion has the final interaction with the system. Without
loss of generality, one can write
\begin{equation}
  \label{eq:matrixelementdef}
  2E_p\frac{dN^{(1)}}{d^3p}=\sum_f\left| 
    \int\! d^4x~ T_f(x) e^{ip\cdot x}\right|^2.
\end{equation}
The definition of the source function for pions is
\begin{equation}
\label{eq:sourcefuncdef}
  S(p,x)\equiv \sum_{f} \int\! d^4\delta x~
  T^*_f(x+\delta x/2) T_f(x-\delta x/2) e^{-ip\cdot \delta x},
\end{equation}
which leads to the simple relation
\begin{equation}
  2E_p\frac{dN^{(1)}}{d^3p}=\left.  \int\! d^4x~ S(p,x) \right|_{p_0=E_p}.
\end{equation}
The source function can be interpreted as the probability per unit space-time
for creating a pion of momentum ${\bf p}$. 

Since source functions can be extracted from semiclassical simulations or
thermal models, it has proven useful to also express two-particle probabilities
in terms of source functions. The two-particle probability requires a
two-particle matrix element $T^{(2)}_f(x_a,x_b)$. Assuming independent, or
uncorrelated, emission means that the two-particle element factorizes,
\cite{Bertsch:1993nx}.
\begin{equation}
  T^{(2)}_f(x_a,x_b)\longrightarrow T_{f_a}(x_a)T_{f_b}(x_b)
\end{equation}
Here, $a$ and $b$ label independent sources. The two-particle probability then
becomes
\begin{eqnarray}
  (2E_1)(2E_2)\frac{dN^{(2)}} {d^3p_1d^3p_2} &=&
  \sum_{a,b,f_a,f_b}
  \left|\int\! d^4x_a d^4x_b~
    T_{f_a}(x_a)T_{f_b}(x_b)
    \frac{1}{\sqrt{2}} \left\{
      \exp(ip_1\cdot x_a+ip_2\cdot x_b)+ \exp(ip_2\cdot x_a+ip_1\cdot x_b)
    \right\}
  \right|^2\\
  &=& \sum_{a,b} \biggl\{
    \int\! d^4x_a d^4x_b ~ S_a(p_1,x_a) S_b(p_2,x_b) \nonumber \\
    && +\int\! d^4x_a d^4x_b~ S_a(\bar{p},x_a) S_b(\bar{p},x_b) 
    \exp[i(p_2-p_1)\cdot(x_a-x_b)]
  \biggr\}\\
  &=& \sum_{a,b} \left\{
    I_a(p_1,p_1)I_b(p_2,p_2)+I_a(p_1,p_2)I_b(p_2,p_1)
  \right\},
\end{eqnarray}
where $\bar{p}=(p_1+p_2)/2$, and the functions $I_{a,b}$ are defined as 
\begin{equation}
  \label{eq:iadef}
  I_{a,b}(p_1,p_2) = \int\! d^4x~
  S_{a,b} \left(\frac{p_1+p_2}{2},x \right) e^{i(p_2-p_1)\cdot x}.
\end{equation}
The above allows one to write the correlation function as
\begin{equation}
  \label{eq:correl_i}
  C({\bf p}_1,{\bf p}_2)=
  1+ \frac{ \sum\limits_{a,b} I_a(p_1,p_2) I_b(p_2,p_1)}
  {\sum\limits_{a,b} I_a(p_1,p_1) I_b(p_2,p_2)}.
\end{equation}
This formulation is similar to that originally used by Shuryak
\cite{Shuryak:1973kq}.

Simulations typically provide a sampling of the on-shell source function. The
application of Eq.~(\ref{eq:iadef}) to simulations is made difficult because
the source functions are evaluated at $\bar{p}_0\ne E_{(p_1+p_2)}/2$ meaning
that they require off-shell information.  The above formalism can be related to
simulations through the smoothness approximation
\cite{Pratt:1997pw,Aichelin:1996iu},
\begin{equation}
\label{eq:smoothness}
  S_a \left(\frac{p_1+p_2}{2},x_a\right) S_b\left(\frac{p_1+p_2}{2},x_b\right)
  \longrightarrow S_a\left(p_1,x_a\right) S_b\left(p_2,x_b\right).
\end{equation}

For thermal sources, one can justify a form for the off-shell behavior of $S$
and the smoothness approximation can be averted. A truly quantum theory would
provide the $T$ matrices that would also allow off-shell evaluation of $S$, and
in fact, formalisms have been developed where classical simulations are
augmented by converting the point particles into wave packets
\cite{wave_packet_formalism}. This also allows one to forego the smoothness
approximation, but at the price of inserting an ansatz for the quantum behavior
that has some peculiarities. We will sidestep the issue of the smoothness
approximation in this study as we wish to focus on quantum aspects associated
with the propagation of off-shell particles.

When calculating correlation functions from simulations, particles from
resonances are usually included in the source function by using the space-time
points from which the resonances decay. The lifetime of the resonance affects
the correlation function through the exponential decay that is simulated in
the transport model. As will be demonstrated later in the paper, the
exponential decay law is modified if the dynamics emit resonances with a
particular mass or range of masses. In this case, the form of the source
function becomes nonexponential as will be explained in the next two sections.

\section{Correlations from resonant sources}
\label{sec:quantum_formalism}

In this section we present a formalism for calculating two-particle correlation
functions from resonance decays given that one or both of the pions might
originate from a resonance. The result will depend on the source function that
represents the creation of the resonance rather than the source function that
represents the points at which the final-state pions are created. The evolution
and decay of the resonance will be accounted for through the quantum propagator
of the resonance. The space-time point at which the resonance decayed to
produce the final-state pions will be treated as an intermediate quantum step
in the evolutionary path between the initial creation of the resonance and the
asymptotic momentum states of the decay products. The expressions derived here
thus incorporate an integration over all points at which the resonance might
have decayed relative to the points at which the resonance is created.

We will consider three examples, a hypothetical scalar $A$ that decays into
two pions, $\rho$, which is a vector resonance also decaying into two pions,
and $\omega$, which is also a vector resonance, but decays into three
pions.
The matrix element for creating the pion and a second particle with momentum
${\bf k}$ through a scalar resonance $A$ is
\begin{equation}
  \label{eq:T_A} T_{\pi,f_A,k}(x)= g\int\! d^4x_A~  \tilde{G}_A(x-x_A)
  T_{A,f_A}(x_A) e^{ik\cdot x},
\end{equation}
where $\tilde G_A$ is the Fourier transform of the propagator for the
resonance and $f_A$ refers to the state of the remainder of the source.
Here $T_\pi$ is effectively the $T$-matrix element for emission of the pion,
while $T_A$ is the $T$-matrix element that would describe emission of the
resonance if the resonance were stable.

Following the same method as in the previous section, one can use $T_\pi$
to create the pion source function using Eq.~(\ref{eq:sourcefuncdef}) to
express $T_A^*(x)T_A(x^\prime)$ in terms of the source function of the
resonance. The resulting expression for the pion source function can be used to
generate $I_A$, as defined in Eq.~(\ref{eq:iadef}), which is all that is needed
to calculate correlation functions,
\begin{eqnarray}
  \label{eq:iascalar}
  I_A(p_1,p_2)
  &=&  g^2\int\! \frac{d^3k}{2E_k}
  \int\! d^4x_A~  \exp[i(p_2-p_1)\cdot x_A]
  S_A\left(\frac{p_1+p_2}{2}+k,x_A\right) 
  G_A(p_1+k) G_A^*(p_2+k) \\
  G_A(p)&=&\frac{i}{p^2-M_A^2+i\Pi_A(p^2)}\\
  \label{eq:scal_propagator}
  \Pi_A(m^2)&=&M_A\Gamma_A\frac{q}{q_R} \frac{M_A}{m}.
\end{eqnarray}
Here, $m^2= p_A^2$, $M_A$ and $\Gamma_A$ are the mass and width of the
resonance, respectively. The relative momentum of the outgoing pions in the
frame of the resonance is $ q^2=m^2/4-m_\pi^2$ and $q_R$ is the same quantity
for an on-shell resonance. We emphasize that the interference term can be
calculated in Eq.~(\ref{eq:iascalar}) without reference to the direct source of
the pions, as the source function of the resonance becomes the required input.

In order to understand the role of the propagators, we reconsider the case of
emitting a pion pair, with momenta $p$ and $k$ through a single, scalar
resonance with momentum $p_A=p+k$,
\begin{eqnarray}
  \label{eq:spectral_explanation}
  2E_p\frac{dN}{d^3p}&=&I_A(p,p)\\
  dN&=&\frac{d^3p}{2E_p} \frac{d^3k}{2E_{k}} d^4x~ S_A(p+k,x)
  \frac{g^2}{\left|(p+k)^2-M_A^2+i\Pi_A\left((p+k)^2\right)\right|^2}\\
  &=&\frac{d^3p_A}{2E_A}d^4x~ S_A(p_A,x)~ dm^2 \frac{1}{\pi}{\rm Im}
  \frac{1}{m^2-M_A^2 -i\Pi_A(p_A^2)}.
\end{eqnarray}
The spectral function describes the density of states of
resonances of invariant masses $m$,
\begin{equation}
  \label{eq:spectral_fct}
  \frac{1}{2\pi}
  {\rm Im}\frac{1}{m^2-M_A^2-i\Pi_A(p_A^2)}=\frac{dn}{dm^2}.
\end{equation}
Thus, one can see that the source function does not provide any information
regarding the mass or width of the resonance. However, combined with the
spectral function, which derives from the product of propagators, it
provides the probability of creating the resonance at space time point $x$ with
momentum $p$ and with invariant mass $m$. For the direct case considered in the
last section, the source function was always evaluated on shell, i.e., the
spectral function was effectively a delta function.

One can also perform the same calculation for vector resonances such as the
$\rho$ meson. In that case, where the coupling of the pions to the vector meson
is
\begin{equation}
  {\cal L}_{\rm int}=i\lambda \left(\pi \partial_\mu\pi\right) \rho^\mu,
\end{equation}
the expression for $I_{\rho}$ is similar to Eq.~(\ref{eq:iascalar})
with the source function and propagator accounting for the vector nature of
the $\rho$,
\begin{equation}
  \label{eq:irho}
  I_{\rho}(p_1,p_2)=\lambda^2\int\! \frac{d^3k}{2E_k}
  \int\! d^4x_\rho~   \exp[i(p_2-p_1)\cdot x_\rho]
  (p_1-k)_\alpha G_{\rho}^{\alpha\beta}(p_1+k) 
  S^{\rho}_{\beta\gamma}\left(\frac{p_1+p_2}{2}+k,x_\rho\right)
  G_{\rho}^{*\gamma\delta}(p_2+k)(p_2-k)_\delta .
\end{equation}
The propagator for the vector resonance is
\begin{equation}
  G_{\rho}^{\alpha\beta}(p)= i\frac{-g^{\alpha\beta}+p^\alpha p^\beta/p^2}
  {p^2-M_A^2+i\Pi_\rho(p^2)}.
\end{equation}
For the vector case the self-energy scales differently  as
a function of the resonance mass than in the scalar case,
\begin{equation}
  \Pi_\rho(m^2)=M_\rho\Gamma_\rho\frac{q^3}{q_R^3}\frac{M_\rho}{m},
\end{equation}
where the same notation as in Eq.~(\ref{eq:scal_propagator}) was used.

As a third example, we consider the propagation of an $\omega$ meson, which is
also a vector resonance, but decays into three pions through 
\begin{equation}
  {\cal L}_{\rm int}=i \kappa \epsilon_{\mu\nu\xi\psi}\omega^\mu
  \partial^\nu\pi^+\partial^\xi\pi^0\partial^\psi\pi^-.
\end{equation}
In this case the expression for $I_{\omega}$ becomes even more complicated
than the $\rho$ example. 
\begin{eqnarray}
  \label{eq:iomega}
  I_{\omega}(p_1,p_2)&=&
  \kappa^2 \int\! \frac{d^3k}{2E_k}\frac{d^3l}{2E_l}
  \int\! d^4x_\omega~ \exp[i(p_2-p_1)\cdot x_\omega] \nonumber\\
  &&\times  \epsilon_{\alpha\mu\nu\xi}~p_1^\mu k^\nu l^\xi
  G_{\omega}^{\alpha\beta}(p_1+k+l) 
  S^{\omega}_{\beta\gamma} \left(\frac{p_1+p_2}{2}+k+l,x_\omega\right)
  G_{\omega}^{*\gamma\delta}(p_2+k+l)
  \epsilon_{\delta\mu^\prime\nu^\prime\xi^\prime}
  ~p_2^{\mu^\prime}k^{\nu^\prime}l^{\xi^\prime}.
\end{eqnarray}
The expression for the self-energy is also somewhat more complicated.
\begin{eqnarray}
  \Pi_\omega(m^2)&=& 
  B \int\! \frac{d^3k}{2E_k} \frac{d^3l}{2E_l}
  ~\delta\left((m-E_k-E_l)^2-E_{{\bf k}+{\bf l}}^2- 2kl\cos\theta \right)
  m^2|{\bf k}\times{\bf l}|^2.
\end{eqnarray}
where $\theta$ is the angle between {\bf k} and {\bf l} and B is an
uninteresting constant fixed by the condition that  
$ \Pi_\omega(M_\omega^2)= M_\omega \Gamma_\omega$, $M_\omega$ and
$\Gamma_\omega$ being the mass and the width of $\omega$, respectively.
After applying the delta function, there remains a fairly complicated
expression, which was evaluated numerically in our considerations of the
$\omega$ case later in the paper.

\section{Limit of a narrow resonance}
\label{sec:narrowresonance}

For this discussion we shall, for simplicity, consider the source function for
a narrow Breit-Wigner resonance $A$. Then, according to
Eqs.~(\ref{eq:matrixelementdef}) and (\ref{eq:T_A}), the probability of
emitting a pion pair with momenta $k$ and $p$ is 
\begin{eqnarray}
  (2E_k)(2E_p)\frac{dN}{d^3p~d^3k}&=& \sum_f\left|
    g \int\! d^4x d^4x_A~ \tilde{G}_A(x-x_A) T_{A,f,k}(x_A) e^{i(p+k)\cdot x}
  \right|^2\\
  &=&g^2 \int\! d^4x_A d^4x~ S_A(p+k,x_A) K(p+k,x-x_A).
\end{eqnarray}
Here, $K$ represents the probability of a resonance carrying momentum $p+k$
propagating from $x_A$ to $x$. It may be expressed in terms of the
propagators,
\begin{equation}
  K(p,x)=\int\! \frac{d^4\delta q}{(2\pi)^4}~ 
  e^{i\delta q\cdot x}G_A^*(p+\delta q/2)G_A(p-\delta q/2).
\end{equation}

In general $K$ is a complicated function. In order to illustrate the quantum
nature, we consider the limit of a narrow resonance that allows the
propagator to be expressed in a simplified form,
\begin{equation}
  G_A(p)=\frac{1}{2M_A}\frac{i}{p_0-E_p+i\Gamma/2}.
\end{equation}
Furthermore, we consider the case where the particles are emitted with equal
and opposite momentum. Integrating over spatial coordinates gives the
probability of the resonance propagating for a time $t$ with off-shellness
$\Delta E$,
\begin{equation}
  {\cal R}(\Delta E,t)
  \equiv \frac{2\Gamma M_A^2}{\pi} \int\! d^3x~ K(p,x)
  = \frac{\Gamma}{\pi}\Theta(t)
  e^{-\Gamma t}\frac{\sin(2\Delta Et)}{\Delta E},
\end{equation}
where $\Delta E=E_k+E_p-M_A$ is the off-shellness. If one integrates over the
off-shellness, the expected exponential behavior is obtained.
\begin{equation}
  \int\! d(\Delta E) ~{\cal R}(\Delta E,t)=\Gamma e^{-\Gamma t},
\end{equation}
whereas integrating over $t$ describes the preference for emitting the particle
with energy close to on-shell.
\begin{equation}
  \int\! dt ~{\cal R}(\Delta E,t)
  =\frac{1}{\pi}\frac{\Gamma/2}{\Delta E^2+(\Gamma/2)^2}.
\end{equation}

The oscillating term $\sin (2\Delta E t)$, which is responsible for
preferentially emitting resonances with small $\Delta E$, also governs the
distribution of emission times. Classical simulations, which are typically
based on Monte Carlo algorithms, cannot easily incorporate regions with
negative probabilities as suggested by this form. The mean propagation time in
a transport simulation could be altered to match the mean time of the quantum
propagator,
\begin{equation}
  \label{eq:NRspectral_fct}
  \langle t\rangle=\frac
  { \int\! dt~t~{\cal R}(\Delta E,t)}
  { \int\! dt~{\cal R}(\Delta E,t)}=\frac{\Gamma/2}{\Delta E^2+(\Gamma/2)^2}.
\end{equation}
However, the second moment for the time would not correspond to that expected
for an exponential decay with the same average time.
\begin{equation}
  \langle t^2\rangle=\frac{1}{8}
  \frac{3\Gamma^2- 4\Delta E^2} { \left( \Delta E^2 + \Gamma^2/4 \right)^2 }
  \ne 2\langle t\rangle^2.
\end{equation}
Not only is this form inconsistent with exponential decay, $\langle t^2\rangle$
can even become negative for resonances far off shell.  This illustrates, on a
formal level, the need for performing the quantum corrections described in this
paper. Quantitative comparisons follow in the next section.

\section{Correlation weights}
\label{sec:weights}

Simulations of heavy-ion collisions usually provide the creation points of
pions along with their outgoing momenta. Neglecting other interactions besides
symmetrization, correlation weights for pions of momenta $p_1$ and $p_2$
originating from space-time points $x_a$ and $x_b$ are usually determined by
calculating the average symmetrization weight of all pairs satisfying the
imposed binning or acceptance \cite{pratt_qm93}. 
\begin{eqnarray}
  \label{eq:correl_weights}
  C(p_1,p_2) &=&1+\frac{\sum\limits_{a,b} \int\! d^4x_a d^4x_b~
    S_a(p_1,x_a) S_b(p_2,x_b) w_a(p_1,p_2) w_b(p_2,p_1)}
  {\sum\limits_{a,b} \int\! d^4x_a d^4x_b~
    S_a(p_1,x_a) S_b(p_2,x_b)}\\
  &\equiv& 1+
  \langle w_a(p_1,p_2) w_b(p_2,p_1) \rangle .
\end{eqnarray}
Inspection of Eqs.~(\ref{eq:iadef}), (\ref{eq:correl_i}), and
(\ref{eq:smoothness}) reveals the weight for direct sources,
\begin{eqnarray}
  \label{eq:weight}
  w_{{\rm d}} (p_1,p_2)&=&\exp[i(p_1-p_2)\cdot x_d]
  \frac{S(\frac{p_1+p_2}{2},x_d)}{S(p_1,x_d)}\\
  w^{({\rm sc})}_{{\rm d}}(p_1,p_2) &=& \exp[i(p_1-p_2)\cdot x_d],
\end{eqnarray}
where the semiclassical form assumes the smoothness approximation
\cite{Pratt:1997pw,Aichelin:1996iu}. 

If the particle originates from the decay of a scalar resonance $A$, the weight
takes on a different form as can be seen from inspecting
Eq. (\ref{eq:iascalar}),
\begin{equation}
  w_A (p_1,p_2) =
  \exp[i(p_1-p_2)\cdot x_A]
  \frac{S_A(\frac{p_1+p_2}{2}+k,x_A)} {S_A(p_1+k,x_A)}
  \frac{(p_1+k)^2-M_A^2 -i\Pi_A[(p_1+k)^2]}
  {(p_2+k)^2-M_A^2 -i\Pi_A[(p_2+k)^2]}.
\end{equation}
Here the resonance was created at $x_A$ and decayed into pions of momenta $p_1$
and $k$. The space-time coordinate of the decay does not enter the weight as
all decay points have been considered.

If the decay is of a vector resonance such as a $\rho$ meson, the weights are
slightly different,
\begin{eqnarray}
  w_\rho (p_1,p_2) &=&
  \exp[i(p_1-p_2)\cdot x_\rho]
  \frac{S_\rho(p_1,p_2,k)}{S_\rho(p_1,p_1,k)}
  \frac{(p_1+k)^2-M_\rho^2 -i\Pi_\rho[(p_1+k)^2]}
  {(p_2+k)^2-M_\rho^2 -i\Pi_\rho[(p_2+k)^2]},\\
  S_\rho\left(p_1,p_2,k, x_\rho\right) &=&
  (p_1-k)^\alpha 
  S^{\rho}_{\alpha\beta}\left(\frac{p_1+p_2}{2}+k,x_\rho\right)
  (p_2-k)^\beta.
\end{eqnarray}
In the derivation of $S_\rho$ we have assumed the two pions involved in the
resonance decay to have equal mass.

Any sort of resonance can be included in this manner, including resonances
that decay into three or more bodies. One such example is $\omega$, which
decays into three pions, one of each species. Labeling the momenta of the two
pions, whose symmetrization we ignore, as $k$ and $l$, the weights turn out to
be 
\begin{eqnarray}
  w_\omega (p_1,p_2) &=&
  \exp[i(p_1-p_2)\cdot x_\omega]
  \frac{S_\omega(p_1,p_2,k,l)}{S_\omega(p_1,p_1,k,l)}
  \frac{(p_1+k+l)^2-M_\omega^2 -i\Pi_\omega[(p_1+k+l)^2]}
  {(p_2+k+l)^2-M_\omega^2 -i\Pi_\omega[(p_2+k+l)^2]},\\
  S_\omega(p_1,p_2,k,l,x_\omega)&=&
  \epsilon^\alpha_{\ \mu\nu\xi}~p_1^\mu k^\nu l^\xi
  S^{\omega}_{\alpha\beta} \left(\frac{p_1+p_2}{2}+k+l,x_\omega\right)
  \epsilon^\beta_{\ \mu^\prime\nu^\prime\xi^\prime}
  ~p_2^{\mu^\prime} k^{\nu^\prime} l^{\xi^\prime}.
\end{eqnarray}

Thus, weights can be used to calculate correlation functions for the decay of
any resonance in a rather straightforward manner. The formalism coherently
accounts for all points at which the resonance may have decayed, but requires
information regarding the points at which the resonances were created as well
as information about the accompanying particles in the decay. The only
difficulty comes in assigning the ratios of the source functions, i.e., the
smoothness problem.

For a thermal source, $S_A(p,x)\sim p_0e^{-p_0/T}$ in the scalar case and
$S^{\rho}_{\alpha\beta}(p,x)\sim p_0 e^{-p_0/T}(-g_{\alpha\beta}+p_\alpha
p_\beta/p^2)$ in the vector case. In the thermal cases the Boltzmann factor
cancels out of the ratios used to calculate weights. Thus for the thermal
example, the weights become a product of four factors, a phase arising from the
points at which the resonance $R \ (=A, \rho, \omega)$ is created, a ratio of
energies, a spin factor, and a ratio of propagator denominators.
\begin{equation}
  \label{eq:thermalweight}
  w_R^{({th})} (p_1,p_2) =
  \exp[i(p_1-p_2)\cdot x_R]
  \frac{(\frac{p_1+p_2}{2}+ k^\prime)\cdot n} {(p_1+ k^\prime)\cdot n}
  \frac{\chi_R(p_1,p_2)}{\chi_R(p_1,p_1)}
  \frac{(p_1+ k^\prime)^2-M_R^2 -i\Pi_R[(p_1+ k^\prime)^2]}
  {(p_2+ k^\prime)^2-M_R^2 -i\Pi_R[(p_2+ k^\prime)^2]},
\end{equation}
where $n$ refers to the frame of the thermal source and $k^\prime$ is equal to
either $k$ for a two-body decay, or to $k+l$ for a three-body decay.

The spin factors $\chi_R$ depend on the sort of resonance being considered.
\begin{eqnarray}
  \label{eq:spinfactor_scal}
  \chi_A(p_1,p_2) &=&1\\
  \label{eq:spinfactor_rho}
  \chi_{\rho}(p_1,p_2)&=&
  (p_1-k)^\alpha
  \left(-g_{\alpha\beta}+\bar{p}_\alpha\bar{p}_\beta/\bar{p}^2\right)
  (p_2-k)^\beta,\\
  \label{eq:spinfactor_omega}
  \chi_{\omega}(p_1,p_2)&=&
  \epsilon^\alpha_{\ \mu\nu\xi}~p_1^\mu k^\nu l^\xi
  \left(-g_{\alpha\beta}+\bar{p}_\alpha\bar{p}_\beta/\bar{p}^2\right)
  \epsilon^\beta_{\ \mu^\prime\nu^\prime\xi^\prime}
  ~p_2^{\mu^\prime}k^{\nu^\prime} l^{\xi^\prime},
\end{eqnarray}
where $\bar{p}= (p_1+p_2)/2+ k^\prime$.

However, one cannot easily apply this technique to nonthermal sources because
the off-shell behavior of the source function is not always known. This problem
also confronts calculations with direct sources, and forces one to either
invoke the smoothness approximation, or assume some form for the off-shell
behavior of the source function.

In analogy to the smoothness approximation one might use the thermal weight
$w_R^{({th})}$ assuming a reference frame $n^\mu$ or
simply neglect the ratio of energies in Eq.~(\ref{eq:thermalweight}). In
fact, the ratio $\chi_R(p_1,p_2)/\chi_R(p_1,p_1)$ can also be neglected to
a reasonable approximation as the structure of the correlation function
derives largely from the last factor in Eq.~(\ref{eq:thermalweight}), the
ratio of propagator denominators.

\section{Comparison with semiclassical models}
\label{sec:results}

In this section we compare calculations of the quantum type described in
Sec.~\ref{sec:quantum_formalism} with semiclassical calculations where the
resonance is assumed to propagate classically and decay according to an
exponential form $\exp(-t/\tau)$. For the purposes of this comparison we
choose to model two simplified systems, one of decaying $\rho$ resonances and a
second of decaying $\omega$'s. For each case resonances are produced and
decayed 
with a Monte Carlo procedure according to a thermal distribution characterized
by a temperature of 150~MeV. The mass of $\rho$ is 770~MeV and the width is
chosen to be 150~MeV, while the mass of $\omega$ is 783~MeV and the width
is 8.4~MeV.

To model the uncorrelated emission of pion pairs, we thermally create particles
of momentum $k$ (and $l$ for the $\omega$) by Monte Carlo, then add to the
particles the weight 
\begin{equation}
  W_{i} =
  \frac{E_k\{+E_l\}+E_i}{E_k\{E_l\}E_i}
  \frac{\chi_R(p_i,p_i)}
  {\left|(p_i+k^\prime)^2-M_R^2+i\Pi[(p_i+k^\prime)^2]\right|^2},
\end{equation}
where $i= 1,2$ and the braces indicate terms that appear for $\omega$
only. This weight accounts for the spectral function of the resonances as
described in Eq.~(\ref{eq:spectral_fct}) and for the spin factors $\chi_R$ as
in Eqs.~(\ref{eq:spinfactor_scal})-(\ref{eq:spinfactor_omega}). We note that
the 
weights $W_{1,2}$ are merely used to generate resonances and their products,
and are not related to the correlation weights. If these weights were included
through a keep/reject prescription, they would not need to appear in any of the
following expressions.

Once the pions are statistically generated, one simply calculates the average
weights described previously to generate the correlation functions.
\begin{equation}
  C(p_1,p_2)= 1+ \Re\left(
  \frac{\sum\limits_a W^a_{1} w_R^{(th),a}(p_1,p_2)}{\sum\limits_a W^a_{1}}
  \frac{\sum\limits_b W^b_{2} w_R^{(th),b}(p_2,p_1)}{\sum\limits_b
    W^b_{2}}\right). 
\end{equation}
For our comparison $w_R$ is either the weight for $\rho$, $w_\rho$, or the one
for $\omega$, $w_\omega$. Note that $a$ and $b$ label individual resonances.

In the semiclassical descriptions, the weights are determined by calculating
the expectation value
\begin{eqnarray}
  \label{eq:weight_A_sc}
  w^{({\rm sc})}_R(p_1,p_2,x_R)&=& \langle e^{i(p_1-p_2)\cdot x} \rangle\\
  &=&\exp[i(p_1-p_2)\cdot x_R]\int\! d^4(x-x_R)~
  \delta^3\left[{\bf x}-{\bf x}_R-{\bf v}_R(t-t_R)\right]
  \exp[i(p_1-p_2)\cdot(x-x_R)] \nonumber\\
  &&\times
  \frac{1}{\gamma_R\tau}\exp[-(t-t_R)/(\gamma_R\tau)]
  \Theta(t-t_R)\\
  &=&\exp[i(p_1-p_2)\cdot x_R]\frac{m_R/\tau_R}{m_R/\tau_R+ip_R\cdot(p_1-p_2)},
\end{eqnarray}
where $\gamma_R$ is the Lorentz factor due to the motion of the
resonance. Here, $w^{({\rm sc})}$ assumes an exponential form for the pion
emission, which is characterized by a lifetime $\tau$. The same form for
exponential decays was developed by Padula and Gyulassy \cite{Padula:1989}.

Based on one's perspective, one might choose any of several prescriptions for
the energy dependence of the lifetime $\tau(m)$. We investigate three
possibilities: (1) The lifetime is chosen such that $m/\tau=\Pi(m^2)$. This
choice would be motivated by the form of the propagator.  (2) The lifetime is
chosen to correspond to the average emission time as described in
Sec.~\ref{sec:narrowresonance}, except that the relativistic generalization of
Eq.~(\ref{eq:NRspectral_fct}) is used $\tau = 2m ~\Im (m^2-M_R^2-i\Pi)^{-1}$.
(3) A fixed lifetime $1/\Gamma$ is used.

If the resonance distribution function in coordinate space is independent from
that in momentum space, the interference term in the correlation function
factorizes into a term stemming from the space-time extent of the resonance
source itself and one that arises from the decay process.
\begin{equation}
  C(p_1,p_2) -1 = \Re \left\{ \langle \exp[i(p_1-p_2)\cdot(x_{A}-x_{B})]\rangle
  \left( C^\prime(p_1,p_2) -1 \right) \right\}
\end{equation}
Here, $x_A$ and $x_B$ refer to points at which the resonances are created, and
$\langle \exp[i(p_1-p_2)\cdot(x_A-x_B)]\rangle$ represents the weighted average
using the product of the source functions as the weight.
\begin{eqnarray}
\label{resweight_eq}
\langle \exp[i(p_1-p_2)\cdot(x_A-x_B)]\rangle&=&\left| J_R(p_1-p_2)\right|^2\\
J_R(p_1-p_2)&=&\frac{\sum\limits_A \int d^4x_A S_A[(p_1+p_2)/2+k_A,x_A] 
\exp[i(p_1-p_2)\cdot x_A]}
{\sum\limits_A \int d^4x_A S_A[(p_1+p_2)/2+k_A,x_A]}
\end{eqnarray}
The reduced correlation function, $C^\prime(p_1,p_2)-1$, is similar to the
average of the weights $\langle w_a w_b \rangle$ from Sec.~\ref{sec:weights},
only with the factors of $\exp[i(p_1-p_2)\cdot x_R]$ removed from the weights
in 
Eq. (\ref{eq:thermalweight}). Since $C^\prime$ contains all the relevant
information about the decay, we will focus on the reduced correlation function
for our comparisons.

Figures \ref{fig:rho200} and \ref{fig:rho800} display the reduced correlation
function for the cases where $(p_1+p_2)/2=200$~MeV/$c$ and 800 MeV/$c$,
respectively. The upper panel of each figure displays results for the case
where ${\bf p}_1$ is parallel to ${\bf p}_2$ while the lower panel displays the
results for the case where the two momenta are perpendicular. All three
semiclassical results exhibit significant deviations from the quantum
calculations. 

Figures \ref{fig:omega200} and \ref{fig:omega800} display the same information
but for a thermal source of $\omega$ mesons. In this case, the semiclassical
treatments are significantly more accurate. This was expected as the width of
the $\omega$ is much less than the temperature, which allows the spectral
function of the $\omega$ to be sampled evenly. As described in
Sec.~\ref{sec:narrowresonance}, the distribution of decay times become
exponential when evenly averaged over all masses. 

The overall width of the correlation functions in the upper panels of Figs.
\ref{fig:omega200} and \ref{fig:omega800} can be understood by noting that the
correlation's width should be determined by the condition $\Delta E\tau =1$,
where $\Delta E\approx v_R q$.  The width of the correlation function for the
lower panels, where ${\bf p}_1\perp{\bf p}_2$, is more complicated since it is
more sensitive to the spatial movement of the resonance while it decays. The
correlation functions for the $\rho$ in Figs. \ref{fig:rho200} and
\ref{fig:rho800} are even more complicated since they extend to large relative
momenta where $\Delta E$ approaches $q$. Given the complicated kinematics
involved, it is not surprising that the result is sensitive to the exact form
for the semiclassical treatment.

In the case of the $\rho$, none of the three semiclassical prescriptions for
$\tau(m)$ provides a consistently good approximation to the quantum result for
all scenarios shown in Figs. \ref{fig:rho200} and \ref{fig:rho800}. They all
exhibit significant deviations from the quantum result for high relative
momentum of the pion pair, above 200~MeV/$c$. In the case of the $\omega$, where
only small relative momenta, $q < 200$~MeV/$c$, are relevant, prescription (1)
seems to best reproduce the quantum result, as can be seen in Figs.
\ref{fig:omega200} and \ref{fig:omega800}.

It should be emphasized that the correlation from the non-zero extent of the
resonance source function has been factored out in this calculation. The
deviations of the semiclassical results occur for relative momenta of a
hundred~MeV/$c$ or more. In a heavy-ion collision the factor $\langle
\exp[i(p_1-p_2)\cdot(x_A-x_B)]\rangle$ in Eq. (\ref{resweight_eq}) would tend to
zero for relative momenta much greater than 50~MeV/$c$ due to the large spatial
sizes of the emitting regions. Thus, the form of $C^\prime$ becomes irrelevant
for higher relative momenta unless the source sizes are small, e.g.,
$pp$ collisions.

\section{Conclusion}

Our findings imply that the semiclassical treatments work quite well for the
larger sources considered with heavy-ion collisions. However, for smaller
sources, especially when the resonance widths are comparable to the
temperature and resonances are produced far off-shell, the semiclassical
treatments significantly deviate from the quantum result.  This can be linked
to the failure of the usual exponential decay law when off-shell resonances are
involved as was shown in Sec.~\ref{sec:narrowresonance}.

In Sec.~\ref{sec:weights}, we proposed a method to correctly calculate
correlation functions from semiclassical models, exploiting the source
function of the resonance and its creation point in space-time as opposed to
the creation points of the final-state pions. This method could be easily
applied to generate correlation functions from the event histories of
simulations. When using direct pions, the creation points of the final-state
pions provide all the necessary information for creating correlation
functions. By considering the creation points of the resonances that decayed
into the final-state pions, one is able to coherently account for all
space-time points at which the resonance might have decayed by modifying the
prescription for generating correlation weights as described in
Sec.~\ref{sec:weights}.

\begin{figure}
  \epsfxsize=0.75\textwidth
  \centerline{\epsfbox{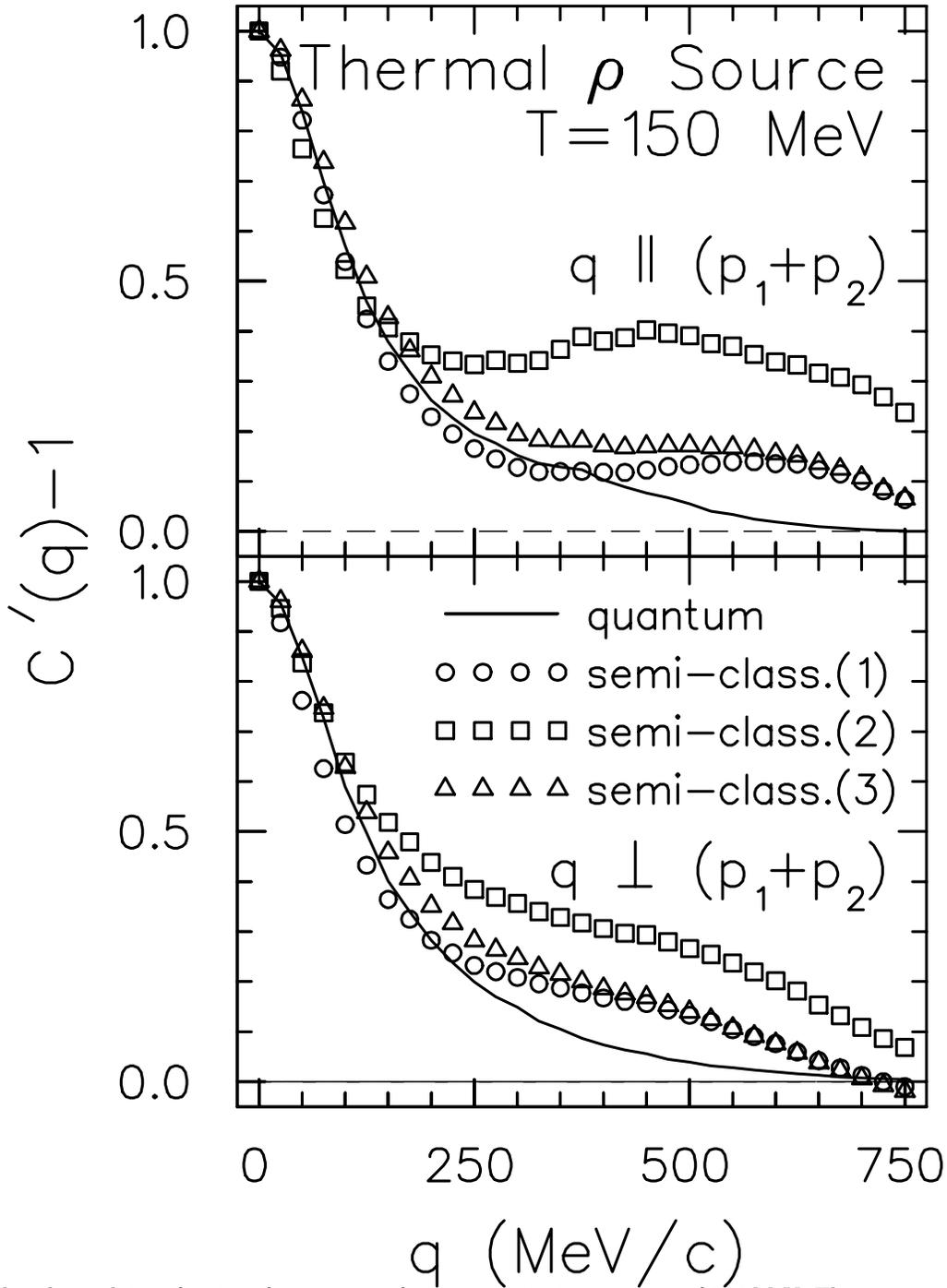}}
  \caption{
    \label{fig:rho200}
    Reduced correlation function for a source of $\rho$ mesons
    at a temperature of 150~MeV. The average momenta of the two pions is fixed
    at $200$~MeV/$c$. By factoring out the space-time dependence of the $\rho$,
    the manifestation of the $\rho$ lifetime is singled out. The exact quantum
    treatment is shown to differ from the three semiclassical treatments
    that are described in the text. The failure of the semiclassical
    descriptions owes itself to the fact that the $\rho$ width is sufficiently
    large for the thermal source to effectively emphasize a specific region of
    off-shellness.} 
\end{figure}

\begin{figure}
  \epsfxsize=0.75\textwidth
  \centerline{\epsfbox{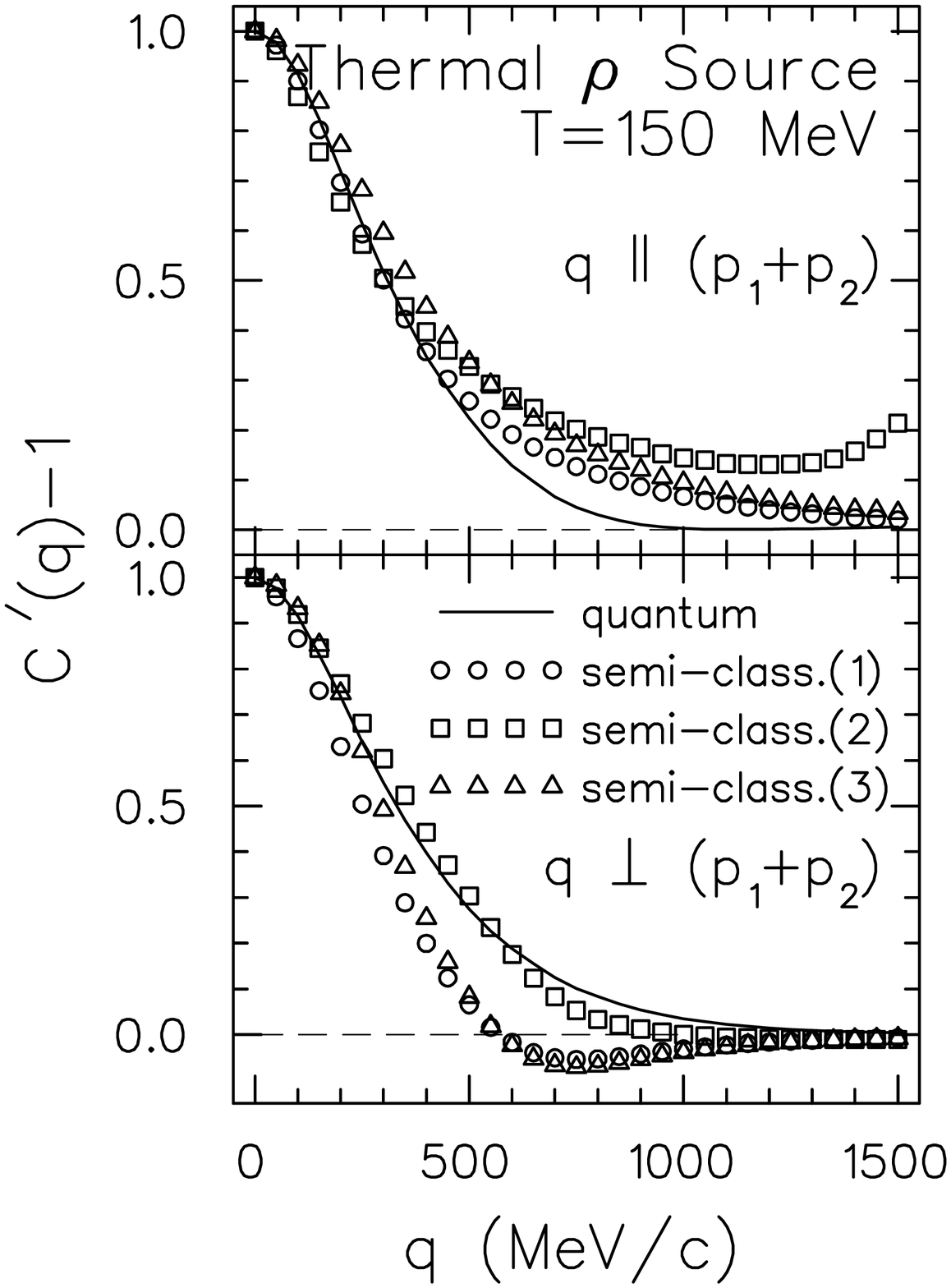}}
  \caption{
    \label{fig:rho800}
    Same as Fig. \ref{fig:rho200}, except that $\rho$'s have average 
momentum $P=800$~MeV/$c$.
    }
\end{figure}

\begin{figure}
  \epsfxsize=0.75\textwidth 
  \centerline{\epsfbox{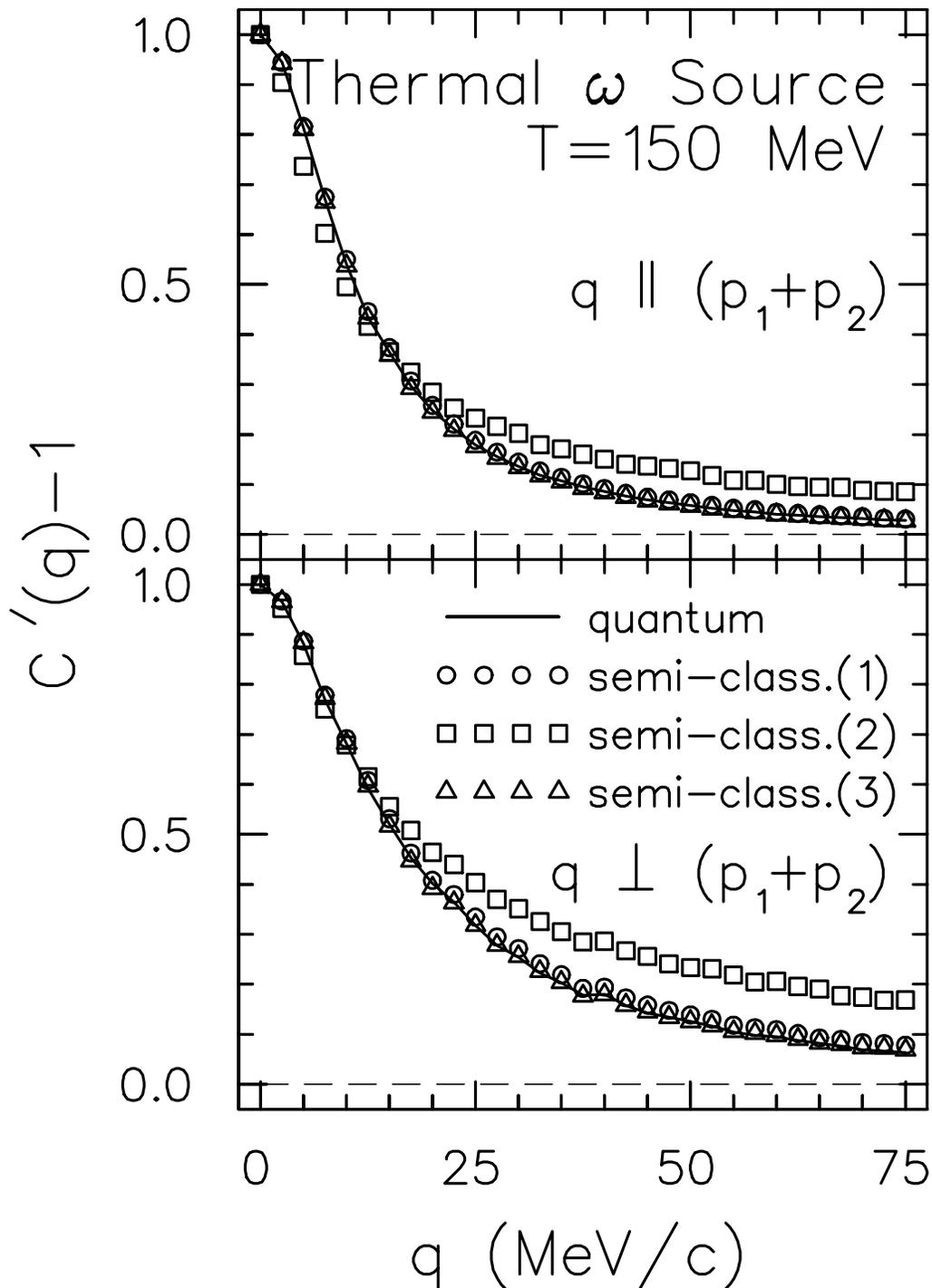}}
  \caption{
    \label{fig:omega200}
    Same as Fig.~\ref{fig:rho200}, but using a thermal source of $\omega$'s
    at a temperature of 150~MeV. In this case, since the width of the $\omega$
    is much less than the temperature, semiclassical treatments work
    remarkably well.}
\end{figure}

\begin{figure}
  \epsfxsize=0.75\textwidth 
  \centerline{\epsfbox{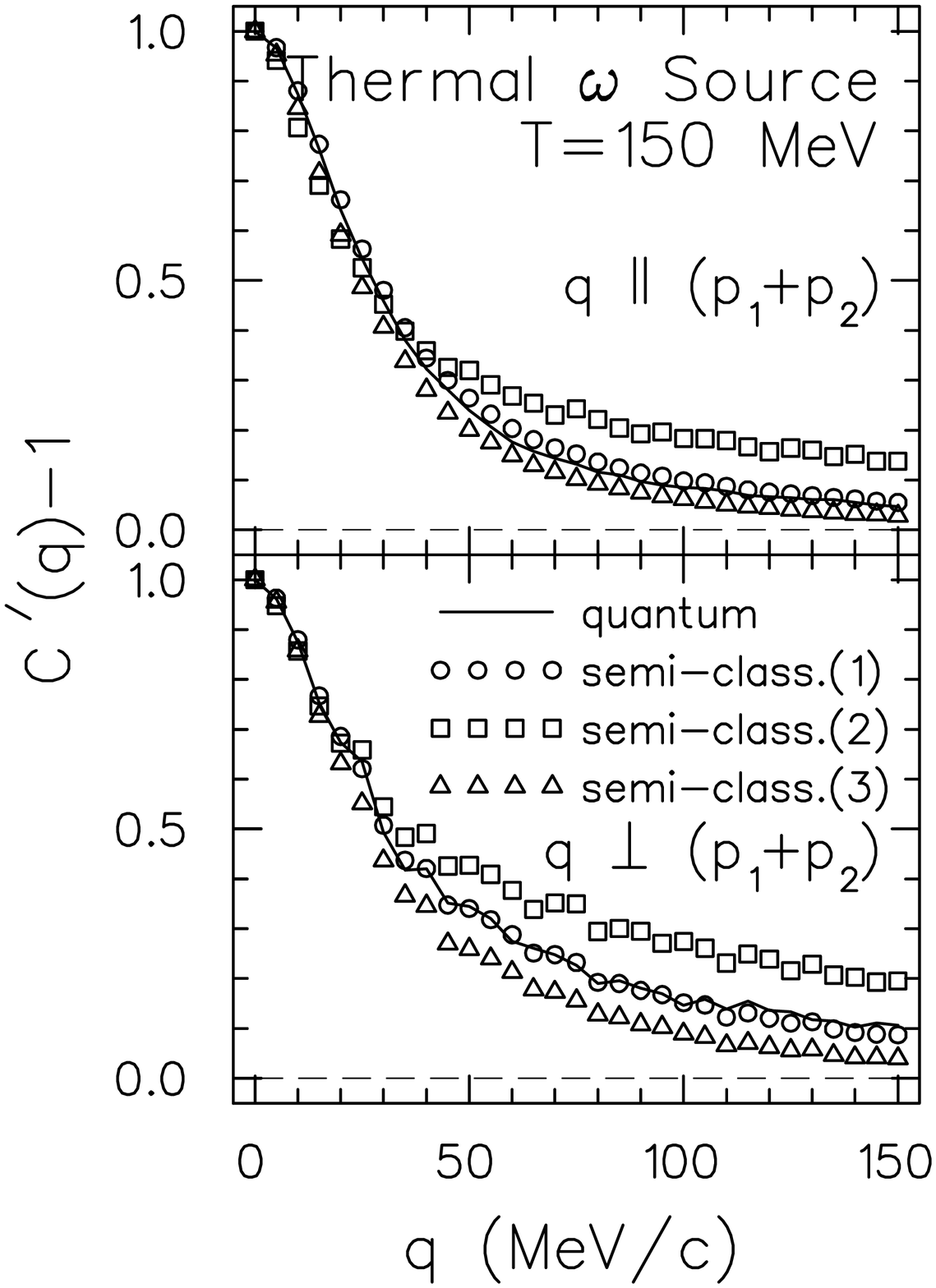}}
  \caption{
    \label{fig:omega800}
    Same as Fig. \ref{fig:omega200}, except that $\omega$'s have average 
    momentum $P=800$~MeV/$c$.
    }
\end{figure}

It should be emphasized that the quantum formalism discussed here becomes
important only for sources that themselves are quantum-mechanical in
nature. If the source is large and the product of the momentum and spatial
uncertainties are large, $\Delta p\ \Delta x \gg \hbar$, the behavior of the
correlation function is dominated by the term $e^{iq\cdot(x_A-x_B)}$, which is
determined by the points at which the resonance is created. Hence $C^\prime$
plays little role in the correlation function of the $\rho$ resonance described
in Fig.~\ref{fig:rho200} when the overall source size is many Fermi as is the
case for heavy-ion collisions. Quantum considerations in resonant decays could
play an important role when considering the decay of small sources that push
the limits of the uncertainty principle. However, such sources are also
accompanied by questions regarding the quantum nature of the source functions
responsible for initial creation of the resonances, i.e., the smoothness
approximation might not be justified. For such problems, unless one knows the
off-shell behavior of the source functions as is the case for a thermal model,
the quantum treatments presented here address only half the problem.

\acknowledgments{
  This work was supported by the National Science Foundation, Grants Nos.
  PHY-96-05207 and PHY-00-70818.}


\begin{thebibliography}{10}

\bibitem{Heinz:1999rw}
U. Heinz and B.~V. Jacak, Annu. Rev. Nucl. Part. Sci. {\bf 49},  529  (1999).

\bibitem{Wiedemann:1999qn}
  U.~A. Wiedemann and U. Heinz, Phys. Rep. {\bf 319},  145  (1999).

\bibitem{pp_gelbke} D.~O. Handzy {\it et al.}, Phys. Rev. Lett. {\bf 75}, 2916
(1995). 

\bibitem{pp_ags} J. Barrette {\it et al.}, Phys. Rev. C {\bf 60}, 054905 (1999).

\bibitem{nn_gaff} S.~J. Gaff {\it et al.}, Phys. Rev. C {\bf 58}, 2161 (1998).

\bibitem{nn_colonna} R. Ghetti {\it et al.}, Phys. Rev. C {\bf 62}, 037603 (2000).

\bibitem{imf_lacey} E. Bauge {\it et al.}, Phys. Rev. Lett. {\bf 70}, 3705 (1993).

\bibitem{imf_desouza} L. Beaulieu {\it et al.}, Phys. Rev. Lett. {\bf 84}, 5971
(2000). 

\bibitem{epluseminus_correlations}O.~Smirnova, B.~L\"orstad, and
  R.~Mure\c{s}an, 
  in {\it Correlations and Fluctuations '98: From QCD to Particle
    Interferometry, Symposia Proceedings},  
  edited by T.~Cs\"org\H{o}, S.~Hegyi, G.~Jancs\'{o}, and R.~C.~Hwa (World
  Scientific, Singapore, 1999), p. 20. 

\bibitem{pratt_qm93} S. Pratt, Nucl. Phys. {\bf A638}, 125
  (1998).

\bibitem{Lednicky:1992me}
R. Lednicky and T.~B. Progulova, Z. Phys. C {\bf 55},  295  (1992).

\bibitem{Bertsch:1993nx}
G.~F. Bertsch, P. Danielewicz, and M. Herrmann, Phys. Rev. C {\bf 49},  442
  (1994).

\bibitem{Shuryak:1973kq}
E.~V.~Shuryak, Phys.\ Lett.\  {\bf 44B}, 387 (1973).

\bibitem{Pratt:1997pw}
S. Pratt, Phys. Rev. C {\bf 56},  1095  (1997).

\bibitem{Aichelin:1996iu}
J. Aichelin, Nucl. Phys. {\bf A617},  510  (1997).

\bibitem{wave_packet_formalism} J. Zimanyi and T. Cs\"org\H{o}, Heavy Ion
Phys. {\bf 9}, 241 (1999);  S.~S. Padula, M. Gyulassy and S. Gavin,
Nucl. Phys. {\bf B329}, 357 (1990).

\bibitem{Padula:1989}
S.~S.~Padula and M.~Gyulassy, Nucl.\ Phys.\  {\bf A498}, 555c (1989).

\end{thebibliography}

\newpage

\end{document}